\documentclass[epj]{svjour}
\usepackage{graphics}
\usepackage{graphicx}
\usepackage{tikz}
\usepackage{csquotes}

\begin{document}
\title{MARTA: A high-energy cosmic-ray detector concept for high-accuracy muon measurement}
\author{
P. Abreu\inst{1,2} \and
S.Andringa\inst{1} \and
P. Assis\inst{1,2} \and
A. Blanco\inst{1} \and
V. Barbosa Martins\inst{7} \and
P Brogueira\inst{2} \and
N. Carolino\inst{1} \and
L. Cazon\inst{1} \and
M. Cerda\inst{3} \and
G. Cernicchiaro\inst{4} \and
R. Colalillo\inst{5} \and
R. Concei\c{c}\~ao\inst{1,2} \and
O Cunha\inst{1} \and
R. M. de Almeida\inst{10} \and
V. de Souza\inst{7} \and
F. Diogo\inst{1} \and
C. Dobrigkeit\inst{8} \and
J. Espadanal\inst{1} \and
C. Espirito-Santo\inst{1} \and
M. Ferreira\inst{1} \and
P. Ferreira\inst{1} \and
P. Fonte\inst{1} \and
U. Giaccari\inst{6} \and
P. Gon\c{c}alves\inst{1,2} \and
F. Guarino\inst{5} \and
O. C. Lippmann\inst{4} \and
L. Lopes\inst{1} \and
R. Luz\inst{1} \and
D. Maurizio\inst{4} \and
F. Marujo\inst{4} \and
P. Mazur\inst{9} \and
L. Mendes\inst{1} \and
A. Pereira\inst{1} \and
M. Pimenta\inst{1,2} \thanks{Corresponding author} \and
R. R. Prado\inst{7} \and
J. \u{R}\'{i}dk\'y\inst{11} \and
R. Sarmento\inst{1} \and
C. Scarso\inst{3} \and
R. Shellard\inst{4} \and
J. Souza\inst{10} \and
B. Tom\'e\inst{1,2} \and
P. Tr\'avn\'{i}\u{c}ek\inst{11} \and
J. V\'{i}cha\inst{11} \and
H. Wolters\inst{1} \and
E. Zas\inst{12}
}                     
\offprints{}          
\institute{
LIP - Laborat\'orio de Instrumenta\c{c}\~ao e F\'{i}sica Experimental de Part\'{i}culas, Braga, Coimbra and Lisbon, Portugal \and
IST - Instituto Superior T\'ecnico, Lisbon, Portugal \and
Observat\'orio Pierre Auger, Malarg\"ue, Argentina \and
CBPF - Centro Brasileiro de Pesquisas F\'isicas, Rio de Janeiro, Brazil \and
INFN - Istituto Nazionale di Fisica Nucleare, Sezione di Napoli and Dipartimento di Fisica \enquote{E. Pancini}, Universit\`a di Napoli \enquote{Federico II}, Naples, Italy \and
Universidade Federal do Rio de Janeiro, Rio de Janeiro, Brazil \and
Universidade de S\~ao Paulo, Instituto de F\'isica de S\~ao Carlos, - IFSC/USP, S\~ao Paulo, Brazil \and
Universidade Estadual de Campinas, IFGW, Campinas, SP, Brazil \and
Fermilab, Chicago, USA \and
Universidade Federal Fluminense, Rio de Janeiro, Brazil \and
Institute of Physics of the Czech Academy of Sciences, Prague, Czech Republic \and
Universidad de Santiago de Compostela, Santiago de Compostela, Spain}
\date{Received: date / Revised version: date}
%
\abstract{
A new concept for the direct measurement of muons in air showers is presented. The concept is based on resistive plate chambers (RPCs), which can directly measure muons with very good space and time resolution. The muon detector is shielded by placing it under another detector able to absorb and measure the electromagnetic component of the showers such as a water-Cherenkov detector, commonly used in air shower arrays. The combination of the two detectors in  a single, compact detector unit provides a unique measurement that opens rich possibilities in the study of air showers. 
%
} 
\maketitle

\section{Introduction}

Over the last decade, the results from the Pierre Auger Observatory \cite{auger} and Telescope Array \cite{ta} have greatly advanced our understanding of the highest-energy cosmic rays. Despite this fact, several fundamental questions remain to be tackled by the upgrades of current experiments or by the next generation of experiments. The determination of the nature of  these highest-energy particles is one of the biggest challenges and an essential ingredient for the astrophysical interpretation of the data \cite{compauger}.

At the highest energies, cosmic rays are scarce and measured indirectly through the detection of extensive air showers (EAS). Large air-shower detectors based only on one technique have limited capabilities for separating the electromagnetic and muonic shower components. The muon component of air showers is rich in valuable information although poorly known. Muons carry information from the first few, high-energy, interactions in the shower. Together with the depth at which the shower reaches its maximum ($X_{\rm max}$), the number of muons in the shower ($N_\mu$) can play a crucial role in the determination of the nature of the primary. To obtain as much information as possible from the showers, disentangling the electromagnetic and muonic components, is paramount for composition studies \cite{lsd}. In particular, a direct and accurate measurement of the muonic component would be a breakthrough.

For the bulk of air showers, information on the muonic component is currently obtained using indirect methods, which lack precision and direct validation. The first measurement of the mean number of muons in inclined ultra-high-energy air showers is presented in \cite{incmu}. The sensitivity of the number of muons to the cosmic-ray mass composition is demonstrated, and a muon deficit in model predictions is observed. To fully explore the constraining power of muon measurements in mass composition, the apparent muon deficit in air shower simulations needs to be understood and the uncertainty on the muon measurement further reduced. Improvements in the description of the muonic component will also reduce the systematic uncertainty in the simulation of the other shower components. 

Measurements presented in \cite{glen} further support that the observed muonic signal is significantly larger than predicted by models. It should be noted that the shower muon content is not being measured directly but instead, these measurements explore shower characteristics to estimate it. It is clear though that current hadronic interaction models cannot provide a consistent description of all shower quantities, signalling deficiencies in their ability to describe the interactions that rule the shower development.
The discrepancy between models and data can only be elucidated by extending this type of event-by-event analysis to include observables with complementary sensitivity to hadronic physics and composition, such as muonic variables, and by improving the ability of detectors to separately measure the muon and electromagnetic components from the signal recorded at ground. Furthermore, studies show that the energy evolution of the moments of $X_{\rm max}$ and $N_\mu$ can be used to assess the validity of a mass composition scenario, surpassing current uncertainties in the shower description \cite{rr}.

In short, current muon measurements are not well described by models and, while this raises interesting questions concerning hadron interaction at the highest energies, large uncertainties come along in the inference of composition. A direct and accurate measurement of the muon component with 100\% duty cycle would allow us to have composition sensitivity on a shower-by-shower basis, to study hadronic interactions at the highest energies, to improve the sensitivity to photon primaries and to better understand and reduce the systematic uncertainties of many different measurements. 

In this paper, we introduce MARTA, an innovative concept combining high-accuracy muon measurements provided by resistive plate chambers (RPCs) with the calorimetric measurement of standard air shower detectors. While MARTA is a generic detector concept, a detailed implementation for the Pierre Auger Observatory has been developed and is used in this work to provide concrete and realistic performance expectations. 

This paper is organized as follows. In section 2, the MARTA concept is presented. The generic detector unit layout and the choice of RPCs are explained. In section 3, details on the built prototypes and performed tests are given. Performances are discussed in section 4. Finally, in section 5, prospects and conclusions are drawn.

\section{Concept and design principles}

\subsection{MARTA concept and design}

MARTA (Muon Array with RPCs for Tagging Air showers) is a hybrid detector concept that combines the information from RPCs with data from another detector able to perform a calorimetric measurement of air showers. The RPCs are placed under the calorimeter, using it as an active shield to most of the electromagnetic shower component, and thus allowing the RPCs to assess directly the muons in the shower. The calorimeter is sensitive both to electromagnetic and muonic components of the shower. As such, it can be used to trigger more efficiently. 


The absorption of the shower electrons and gammas, and thus the purity of the muonic signal,   increases with the quantity of matter over the RPCs. It should be noted, however, that also the energy threshold for muon detection increases with the mass overburden.
Nevertheless, given the typical muon energies, the effect is expected to be small, as discussed below.
Also, a mass overburden of the order of few  
hundreds of ${\rm g\,cm^{-2}}$ is sufficient to grant the accurate reconstruction of the muon component in a wide range of energies and zenith angles. 
The high segmentation of the RPCs allows for the definition of fiducial areas. These fiducial areas, defined selecting RPC pads, are chosen to take into account the effective mass overburden over each individual pad. This quantity depends on the arrival direction of particles, which can be estimated from the reconstructed shower direction, as discussed in the next section.


The hybrid concept of MARTA not only allows for the separate measurement of the shower components, giving additional insight on the shower development mechanisms but also enables better control of systematic uncertainties inherent to each detector through, for instance, cross-calibrations.

\subsection{The Choice of RPCs}

RPCs are widely used in nuclear and particle physics experiments. These gaseous detectors have shown to be robust while having a high particle detection efficiency. Moreover, they are relatively low-cost detectors that can easily offer excellent spatial and time resolutions \cite{rpc}. In the recent years, much R\&D on the ability to run RPCs in outdoor environments and with low maintenance has been performed~\cite{outrpc1,outrpc2}. The results demonstrate that RPCs are a good candidate to be used in cosmic-ray experiments.

RPCs are composed of millimetre-thick gaseous volumes that are contained by highly resistive parallel plates. High-voltage (HV) electrodes are applied to these plates, creating an intense and uniform electric field. The passage of ionising particles through the detector creates avalanches of electrons which induce signals in the readout electrodes. The high resistivity of the plates prevents electrical discharges, which would affect the whole detector.

RPCs have significant advantages with respect to more conventional detectors, as for example scintillators, particularly concerning cost and feasibility. Moreover, the segmentation level is very flexible and constrained essentially by the readout. The signal pick-up electrodes are physically separated from the sensitive volume. This approach allows us to achieve high-voltage insulation and gas tightness, reducing the number of breakthroughs considerably. An aluminium case is used to host the RPC,  the Data Acquisition system (DAQ), the high-voltage and monitoring systems. Details on the design of the assembled and tested prototypes are presented in section 3. The MARTA design is based on a multi-gap gaseous volume. The usage of thin gas gaps guarantees fast detector response to avalanche development, yielding very good time resolutions. Moreover, the multi-gap approach enhances detection efficiency. The chambers require low gas flux and use tetrafluorethane (R-134a), a common refrigerator gas, and the main component of the mixture used in most modern RPC installations. 

The MARTA concept also has advantages when compared to muon detectors buried underground, below an air shower detector. Firstly, the energy threshold for muons remains essentially the same in the MARTA sub-detector (WCD + RPCs), while it would differ considerably between underground detectors and surface ones. The implementation of MARTA would also consume much less time and resources. Finally, as mentioned before, the detection of the same particles by both detectors provides an invaluable tool both to understand the detector performance and to further exploit the shower physics. Other interesting possibilities for combined measurements include primary photon identification and air shower physics near the core.

\section{Implementation and prototypes}

\subsection{A MARTA baseline design}
\label{sec:baseline}
A possible design of MARTA has been elaborated in detail for the Pierre Auger Observatory. Several prototypes have been built and tested both in laboratory and at the observatory site. 

\begin{figure}
\resizebox{0.45\textwidth}{!}{%
  \includegraphics{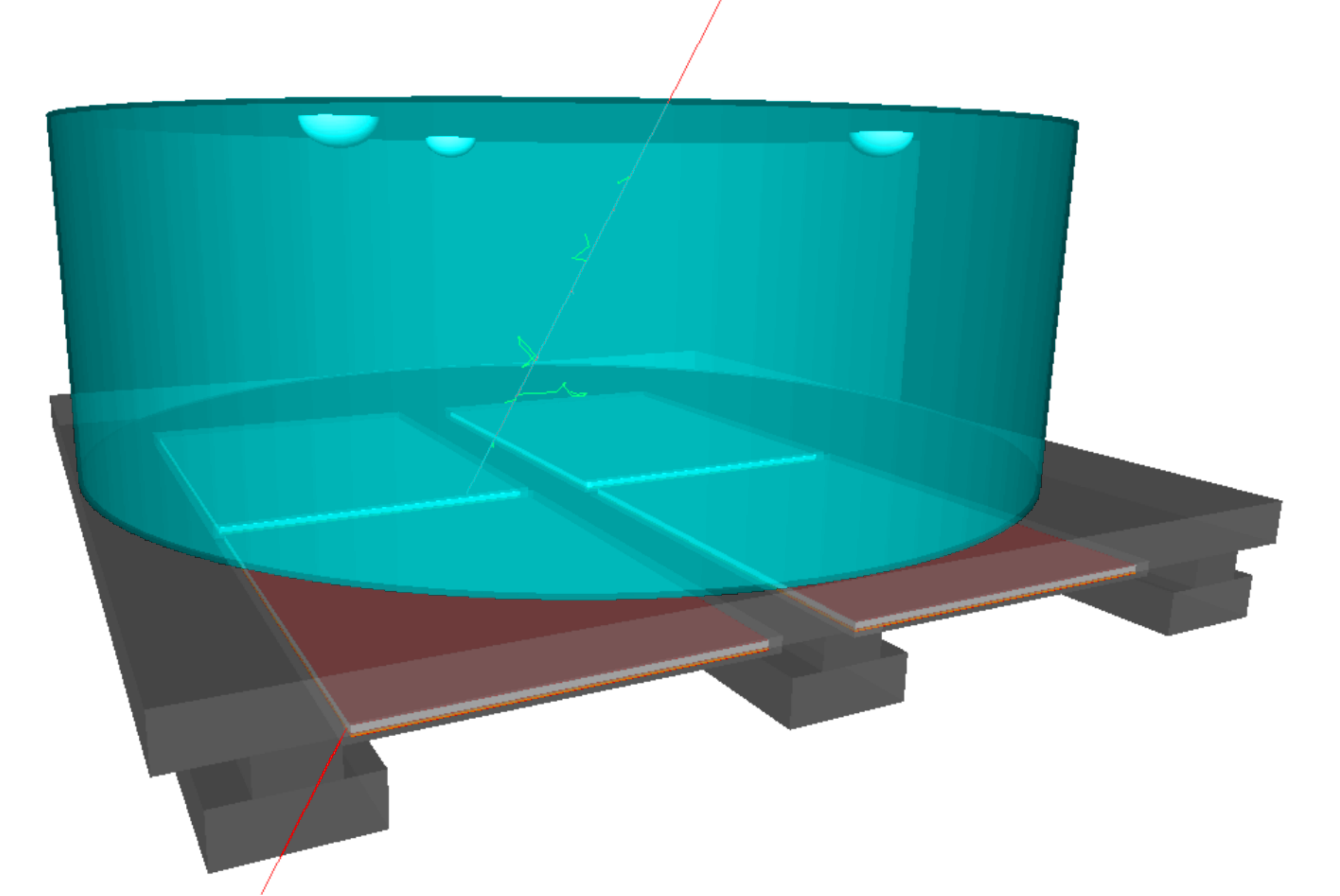}}
\caption{\label{fig:fig3}
The MARTA implementation for the Pierre Auger Observatory: the RPCs (in brown) are placed under the water-Cherenkov detector (in green) which provides active shielding and trigger. The concrete support structure is shown in black.}
\end{figure}

At the Observatory, MARTA units could be installed at the entire surface array under the water-Cherenkov detectors (WCD). The WCD would remain unchanged, acting as shielding for the electromagnetic shower component, and sitting on top of a concrete structure hosting the RPC modules. The water (1.2 m depth) and the concrete (20 cm thickness) correspond to a mass overburden of 170 g/cm$^2$. A schematic view of the MARTA implementation for the Pierre Auger Observatory is displayed i figure \ref{fig:fig3}.

\vspace{2mm}
\noindent\textbf{RPC unit}

The baseline configuration foresees four RPCs per tank. The structure of each chamber is as follows:

\begin{itemize}
\item An area of 1.2 x 1.5 m$^2$, for a total of over 7 m$^2$ of RPC per WCD;
\item A total of three resistive plates made of soda-lime glass, each 2 mm thick, mounted on top of each other;
\item The resistive plates are separated by 1 mm gaps for the gas, making it a double gap chamber, filled with R-134a;
\item The detector is glued to an acrylic box of 3 mm thickness;
\item The readout plane is external and segmented in 8 x 8 pick-up electrodes (pads), each with dimensions  $14\times18\,$cm$^2$ and separated by a 1 cm guard ring;
\item Coaxial cables transmit the signal induced in each pad to the DAQ.
\end{itemize}

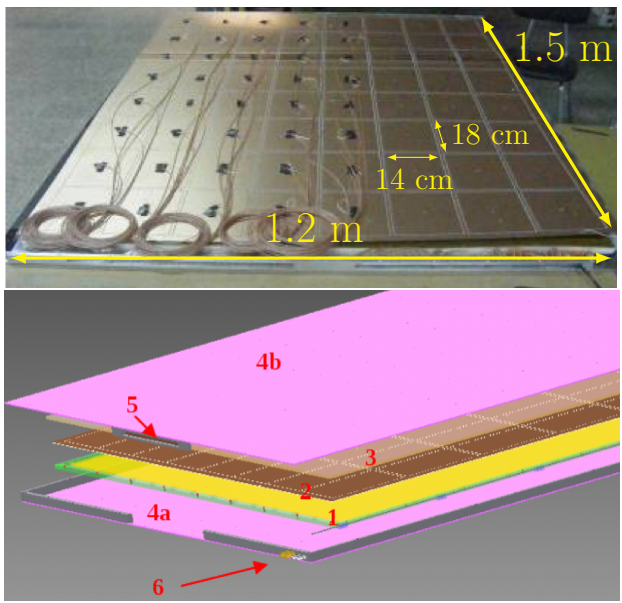
\begin{figure}
\centering
\resizebox{0.45\textwidth}{!}{%
  \begin{tikzpicture}
    \node[anchor=south west,inner sep=0] at (0,0) {\includegraphics[clip,trim= 0cm 1.5cm 0cm 1.25cm]{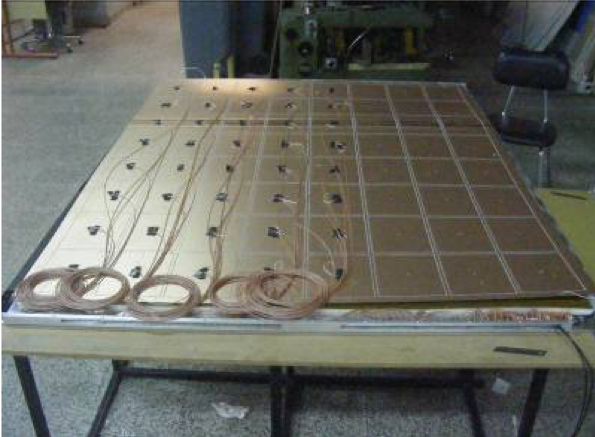}};
    \draw[latex-latex,yellow,line width=1.5pt] (0.1,0.5) -- (10,0.5);
    \node[text=yellow] at (5.1,1) {\huge $1.2\ \mathrm{m}$};
    \draw[latex-latex,yellow,line width=1.5pt] (10,1) -- (7.85,4.5);
    \node[text=yellow] at (9.2,4) {\huge $1.5\ \mathrm{m}$};
    \draw[latex-latex,yellow,line width=.5pt] (6.3,2.17) -- (7.15,2.17);
    \node[text=yellow] at (6.75,1.8) {\Large $14\ \mathrm{cm}$};
    \draw[latex-latex,yellow,line width=.5pt] (7.27,2.25) -- (7.1,2.8);
    \node[text=yellow] at (8,2.55) {\Large $18\ \mathrm{cm}$};
  \end{tikzpicture}}
\resizebox{0.45\textwidth}{!}{%
  \includegraphics[clip,trim= 0.5cm 0.2cm 10cm 3cm,width=0.94\textwidth]{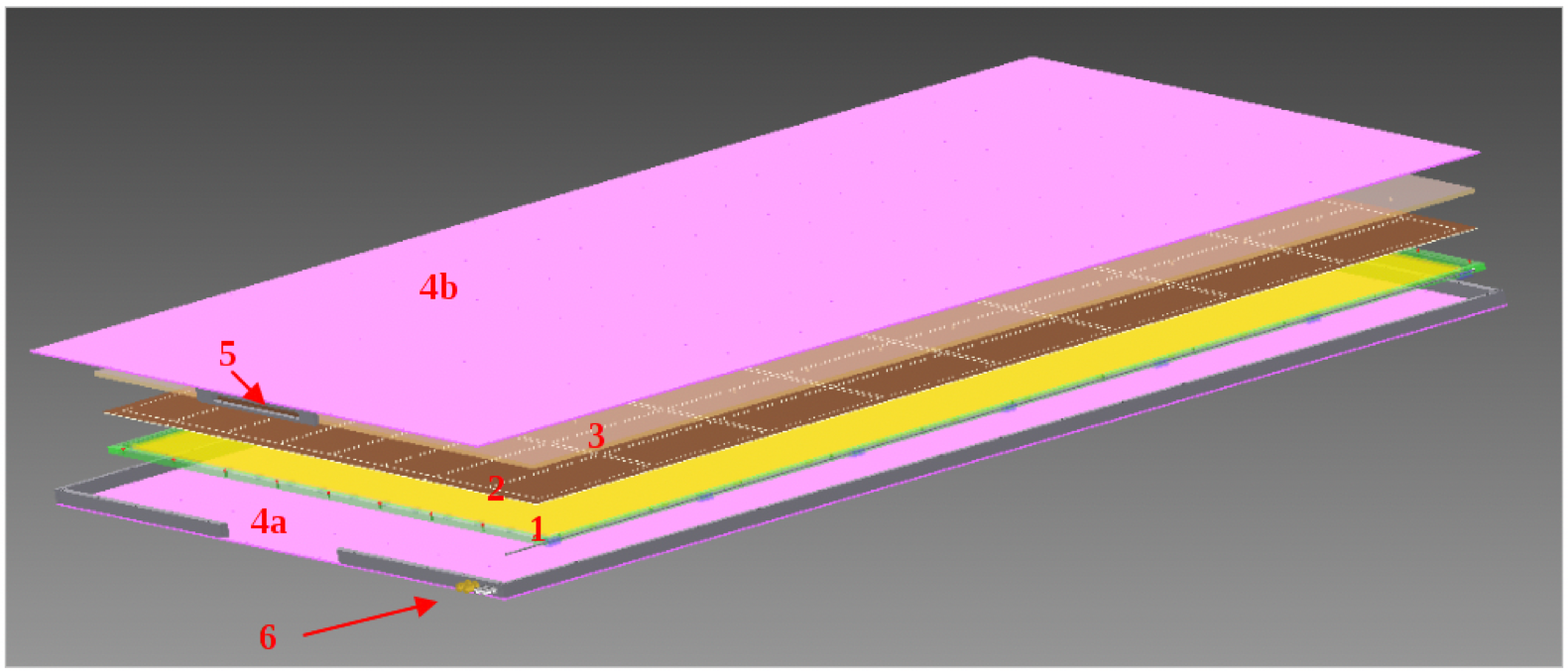}}%
  \caption{\label{fig:fig4}
Photo and scheme of the RPC detector. Top: RPC with visible cabled pad plane showing the detector spacial resolution. Bottom: scheme of the RPC box (1), readout plane (2), I$^2$C sensors layer (3), aluminium case base, cover and junction (4a, 4b, 5), and breakthroughs for gas and high-voltage (6).}
\end{figure}

In figure \ref{fig:fig4}, a photograph and a scheme of the RPC unit are shown. The high-voltage electrode and the active detector layers are enclosed inside the acrylic box, guaranteeing  high-voltage insulation and gas tightness. Only two breakthroughs for the gas and two for the high-voltage are required. The asymmetric design, with the readout electrode at only one side of the gaps, has the advantage of easing the cabling by having it only at one side of the detector.

\vspace{2mm}
\noindent\textbf{Data acquisition system}

A new front-end acquisition system \cite{martadaq} was developed for MARTA RPCs. It is a hybrid system capable of counting active RPC pads and measuring the charge induced in the detector. To comply with the strict demands of MARTA operations in the field, the system was designed to be low-power consuming (a few Watts per RPC), compact (due to space limitations inside the aluminium case), stable and reliable, for low maintenance operation. These requirements made a system based on an application specific integrated circuit (ASIC) the most appealing option. The Multi-Anode Readout Chip, MAROC 3 \cite{maroc}, is a low power (3.5 mW per channel) and compact (16 mm$^2$) ASIC that fulfils all the stated criteria with 64 input channels, 64 discriminated outputs and that can measure charges up to 15 pC.

The ASIC counts particles by applying a simple threshold to the signal after a pre-amplifier and a fast shaper. To measure charge, a slow shaper is applied to the signal after the pre-amplifier. The slow shaper peak is then converted to digital using an analog to digital converter (ADC) and taken as the charge induced in the RPC. Both measurements have been tested~\cite{martadaq}. The results show that the DAQ can measure the fast RPC signals without introducing any unwanted inefficiencies in the setup.

\begin{figure}
  \centering
  \includegraphics[clip, trim=1.9cm 1.75cm 1.95cm 0.75cm,width=0.45\textwidth]{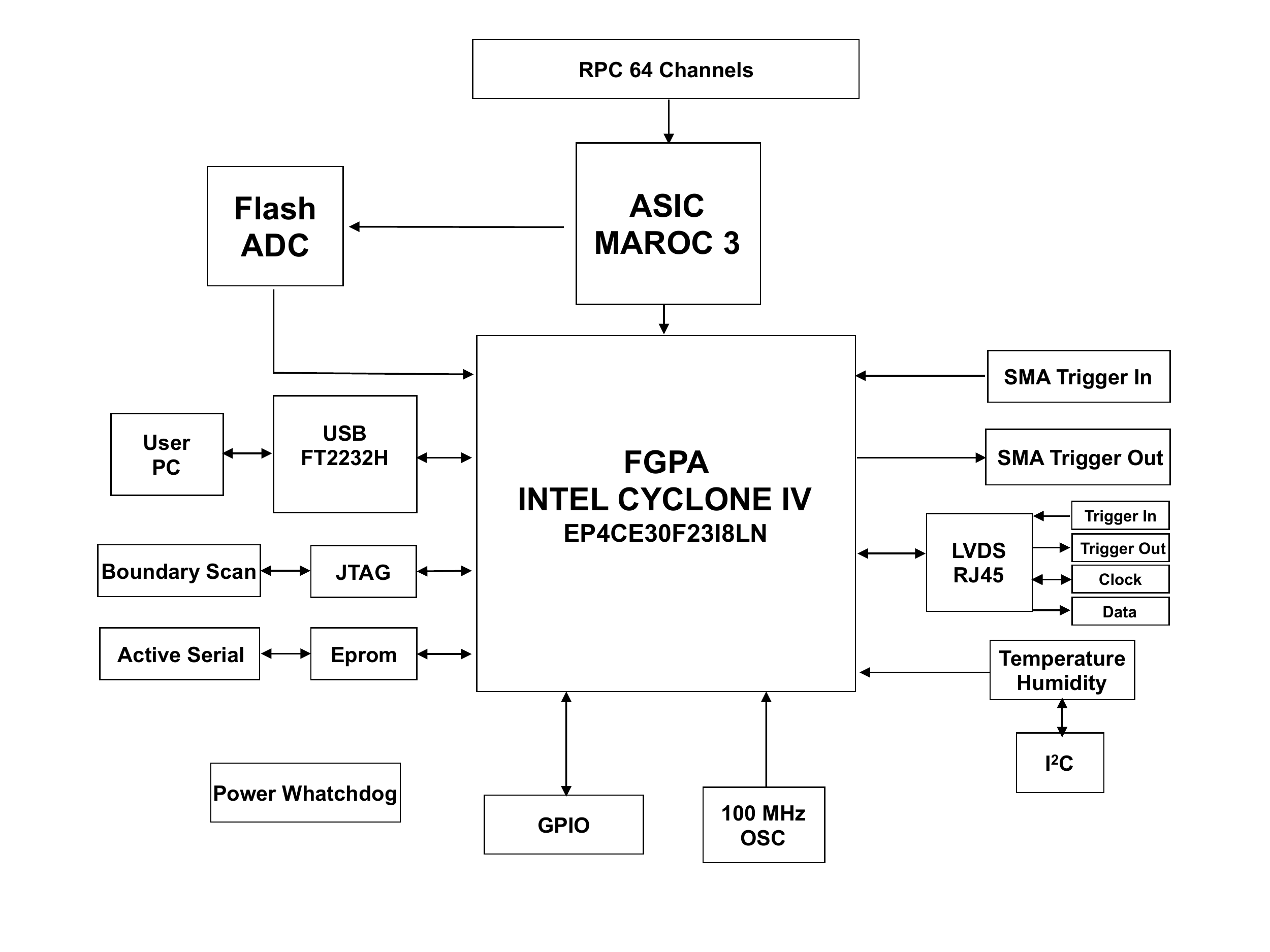}
  \caption{\label{fig:fig5}
MARTA DAQ schematic representation with its main components.}
\end{figure}

A low power field-programmable gate array (FPGA) processes and stores the ASICs digital outputs. It is also responsible for all the communications using low voltage differential signalling (LVDS) lines. These lines, connected to a concentrator central unit, are also used as trigger interface. Alternative communication with a PC via USB is also available and used mostly for debugging. Additional features are available to increase the system flexibility, namely an analog acquisition of the sum of the RPC signals, environmental and power monitoring as well as multi-purpose input and output ports. In figure \ref{fig:fig5}, a scheme of the MARTA DAQ is shown.

\subsection{Prototype construction and test}

Several prototypes of MARTA RPCs were built in the LIP-Coimbra workshops according to the aforementioned specifications, following the requirements for counting individual muons in extensive air showers. These have been tested indoors and outdoors, including at the Pierre Auger Observatory site, with the objective to study the long-term behaviour of the detector, evaluating its resilience to the environmental conditions and monitoring the operational parameters.

The initial laboratory tests were conducted with pure R-134a gas at a very low flow rate, of the order of 0.4 cm$^3$/min or 1 kg/year, with the purpose of evaluating the imperviousness to humidity \cite{outrpc1}. The background current, measured directly from the HV power supply, was chosen as the parameter to monitor the detector conditions, since it depends not only on the gas ionization rate and average charge per ionization, but also on additional contributions from leakage currents. There was no observed increase in the background current with the detector immersed in water for two weeks. Moreover, detector operation was achieved with a fraction of streamers below 10\% and high detection efficiency for cosmic rays. The background current was also monitored during seven months in a chamber placed outdoors, showing no strong dependence on the detector gas pressure or relative humidity.

The indoor tests were prolonged for nine months, using gas flow rates of 1, 4 and 12 cm$^3$/min. A detection efficiency of the order of 90\% was observed in all cases \cite{outrpc3}. This value coincides with the fraction of the detector sensitive area that is covered by the readout pads, meaning an almost 100\% intrinsic detection efficiency for cosmic muons. Moreover, the efficiency dependence on the detector reduced electric field was measured, a curve which is to be used for keeping track of the efficiency under the final measurement conditions. Also, for the first time, some prototypes were installed outdoors at the Pierre Auger
Observatory site, where large daily variations in temperature and pressure occur. Nevertheless, after four months of operation, a temperature variation in the detector (6$^\circ$C) much lower than the ambient thermal amplitude (28$^\circ$C) was observed, which is explained by the thermal inertia of the tank with its concrete support structure, as predicted by a thermal simulation. 

Later software developments allow to dynamically adjust the applied high voltage in function of the average pressure and temperature, to keep a constant value of the reduced electric field. The main results of these developments, reported in \cite{outrpc2} and \cite{outrpc4}, were the confirmation of small daily thermal amplitudes in the detector and a remarkable stability of the efficiency, at the level of 85\%, measured during nearly one year of operation in the field. Constant and uniform efficiency across all the detection area, independent of the temperature or pressure gradients, was also observed. After almost two years of field measurements at the Pierre Auger Observatory, it has been shown that these RPCs can be operated in a harsh outdoor environment, and perform suitably for a cosmic-ray experiment.

During the test phase an application of this detector concept was started: the study of the response of an Auger water-Cherenkov detector (used for tests) to atmospheric muons, by using a hodoscope built with RPC prototypes and custom-made electronics \cite{prec}, with very successful results \cite{gianni}. Currently, a first shower-dedicated measurement is in progress.

\section{Expected performance}

MARTA is a generic detector concept designed to fulfil the requirements of large high-energy cosmic-ray experiments. A detailed implementation for the Pierre Auger Observatory has been developed and is used in this work to provide concrete and realistic performance expectation. A detailed simulation of this implementation of MARTA has been performed using the GEANT4 toolkit \cite{geant4}, according to the baseline design described in section~\ref{sec:baseline}. 
EAS simulations for several primary species, zenith angles and energies were undertaken using CORSIKA~\cite{corsika}. The QGSJet-II.04~\cite{qgsjet} and EPOS-LHC~\cite{epos} have been used as hadronic interaction models.
\subsection{MARTA unit}

\begin{figure}
\resizebox{0.425\textwidth}{!}{%
  \includegraphics{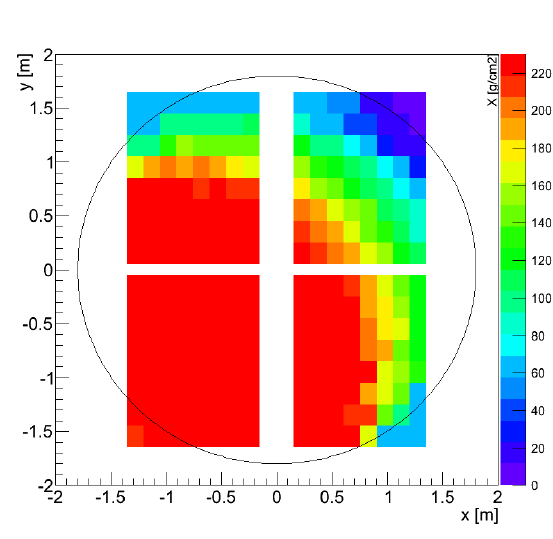}}
\caption{\label{fig:slant-mass}Slant mass crossed before reaching the MARTA RPCs, under a 170 g/cm$^2$ vertical mass overburden, for incident particles at $40^\circ$ zenith angle. The circle indicates the area covered by the water tank.}
\end{figure}

A measurement of the number of muons can be obtained from each individual MARTA detector unit. The first crude estimator of the number of muons is the number of hits in the pads within a fiducial area defined as the set of pads in a given shower that have a mass overburden greater than a chosen value. In the case of the MARTA configuration, the definition of the fiducial area required a slant mass greater than $170\,{\rm g\,cm^{-2}}$, corresponding to the vertical mass overburden from the water tank and the concrete tank support, defining a minimum criteria - with $100\%$ of fiducial volume for vertical events. 
The number of pads within the fiducial area is then a function of the shower geometry. An example of a slant mass map, computed for incident particles at $40^\circ$ zenith angle, is shown in figure~\ref{fig:slant-mass}. In this case 2/3 of the pads are contained in the fiducial area.
For a vertical shower, all the pads located below the shielding detector are contained in the fiducial area.

The dependence of the energy threshold for muon detection with the mass overburden was studied using simulated CORSIKA showers. The muon energy spectrum at 1400~m above sea level peaks above 1~GeV and about 15\% of these muons are absorbed after crossing the additional $170\,{\rm g\,cm^{-2}}$.

\begin{figure}
\resizebox{0.45\textwidth}{!}{%
  \includegraphics{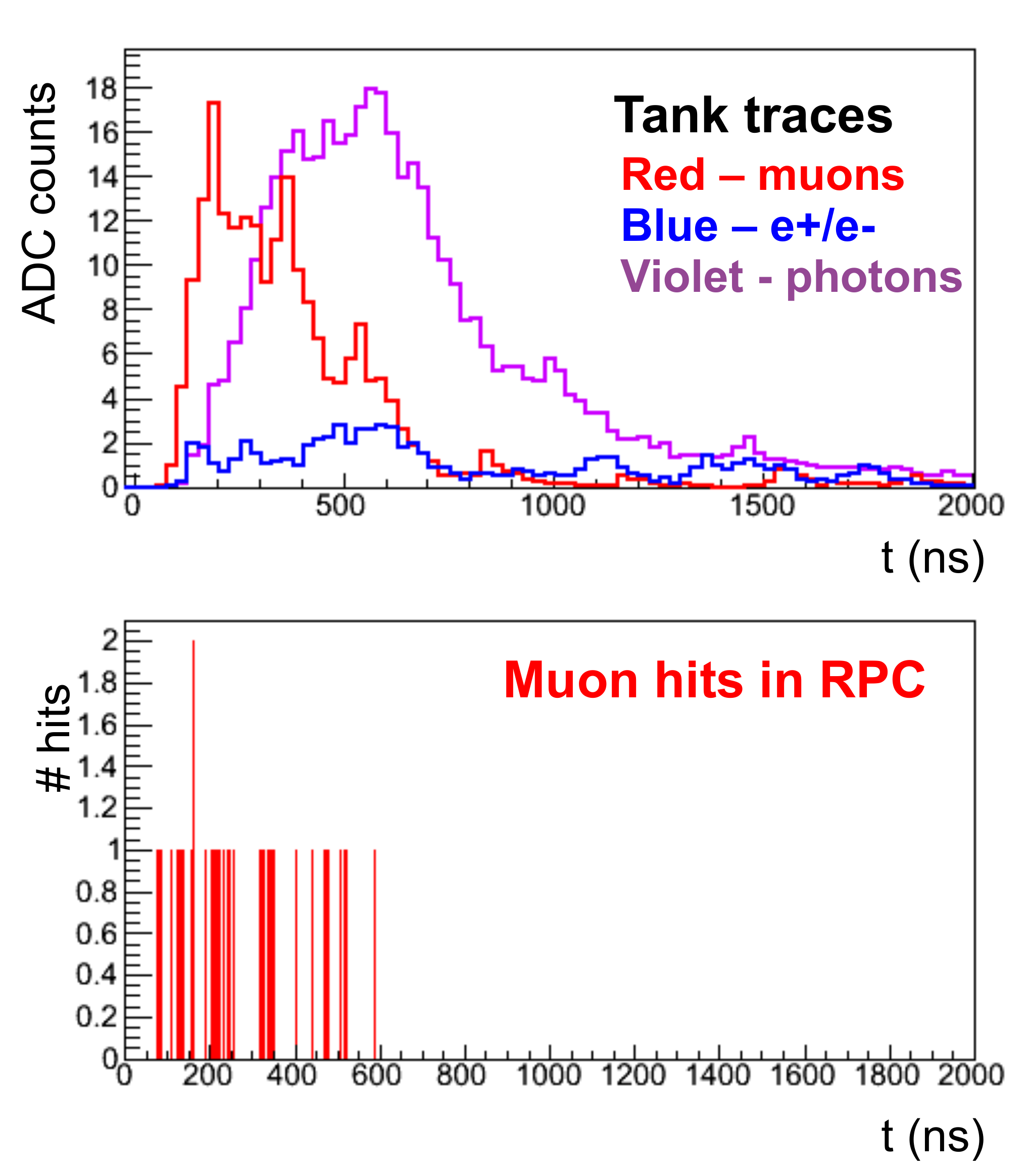}}
\resizebox{0.45\textwidth}{!}{%
  \includegraphics{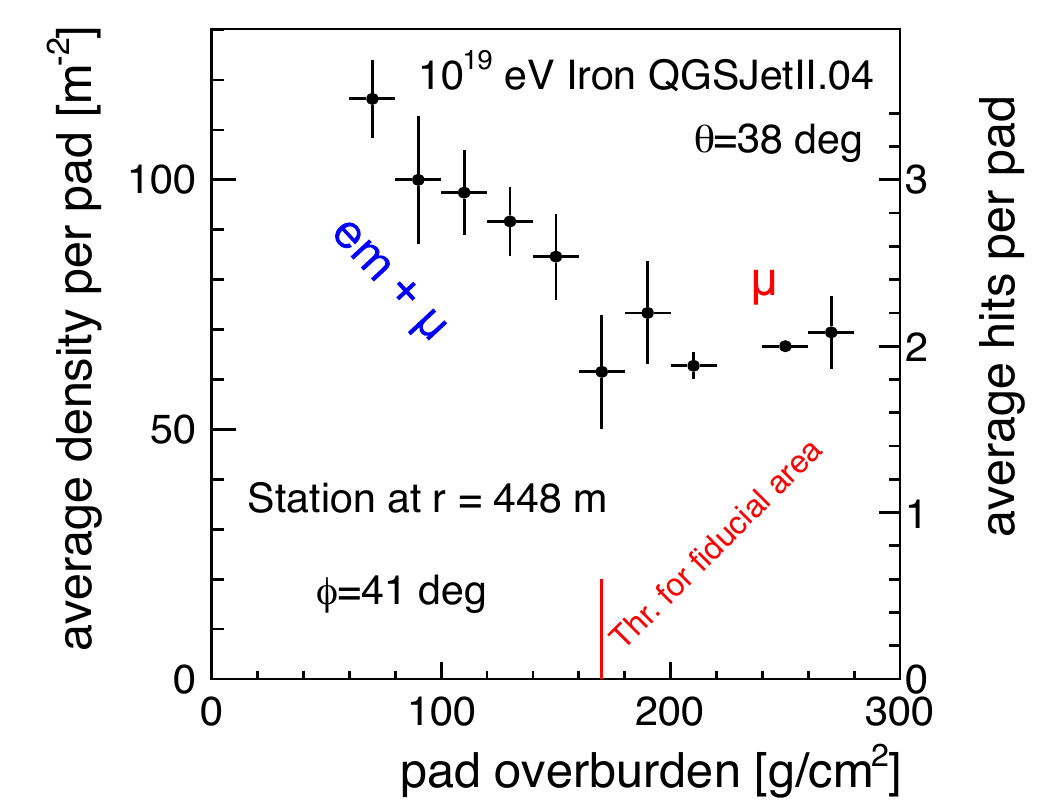}}
\caption{\label{fig:fig6}
Top: Example of a MARTA simulated trace -- the simulated WCD and RPC traces are shown;
Bottom: Separation of the electromagnetic and muonic shower components in a simulated MARTA unit.}
\end{figure}

In figure \ref{fig:fig6}, an example of a trace in the MARTA RPCs is shown and compared with the traces in the water-Cherenkov detector. The muonic signal separation as a function of the pad overburden is also shown. 

The RPC segmentation and the chosen readout electronics allow for the digital counting of muons with a time resolution of 5 ns and a position resolution limited by the pad dimensions. For the baseline design described in section 3, this corresponds to a maximum particle density of 35.6 per m$^2$ (assuming that all particles arrive at the same time). This density is equivalent to that of muons at about 500 m from the shower axis for a proton shower with $E=10^{19.5}\,$eV and $\theta=40^\circ$. Due to the spread of the muon arrival times and the small dead time of the readout, pile-up effects become relevant only at smaller distances and the number of muons can be successfully recovered, in the case exemplified, down to about 300 m by applying dedicated algorithms \cite{supanit}. For the purpose of measuring the signal very near the shower core at the highest energies, the analog channel is expected to provide counting capabilities up to 20000 particles per RPC.

The bias and the resolution of the reconstructed muon signal have been estimated using the digital mode only and no pileup correction. A bias (due to electromagnetic signal contamination) of around 20\% is, as expected, seen down to a distance to the core of 500 m. Below 500 m, the pileup effect starts to be visible and must be corrected for. At $E=10^{19}$ eV, the resolution of the reconstruction of the muonic signal is below 20\% up to 1000 m. For large distances to the shower core, the muon signal resolution is dominated by the low number of secondary particles. 

\begin{figure}
\resizebox{0.45\textwidth}{!}{%
  \includegraphics{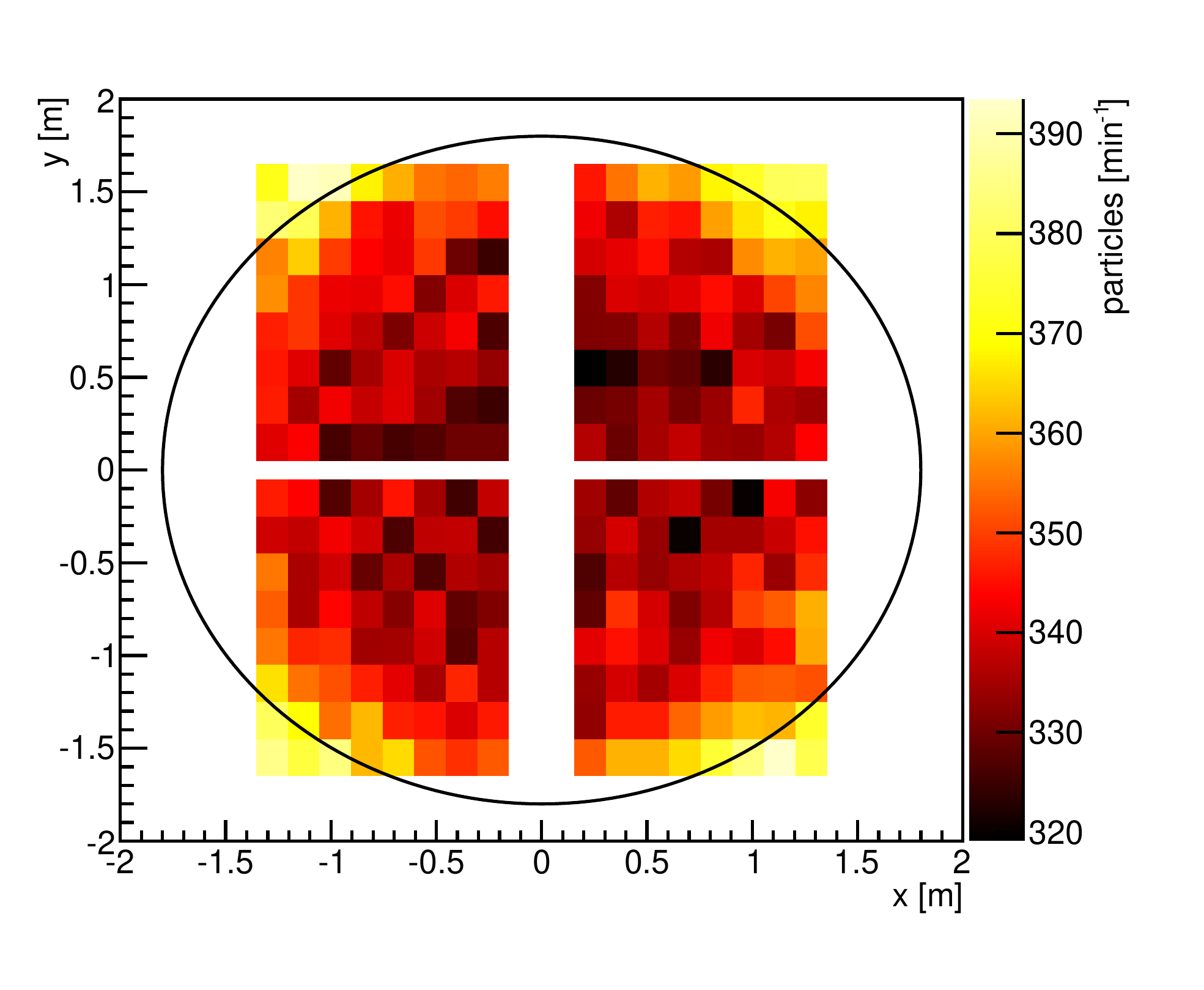}}
\caption{\label{fig:fig7}
Number of atmospheric particles in each RPC pad per minute. Result obtained from a dedicated simulation at $1400\,$m a.s.l. (see text for details).}
\end{figure}

The atmospheric muon flux can be used to monitor and calibrate the efficiency of the pads. Estimations of background muons per pad were done using dedicated simulations. The atmospheric particle flux at ground (1400 m a.s.l.) was obtained through CORSIKA simulations. These simulations were obtained injecting the particles according to the known primary cosmic ray energy spectrum, with energies ranging from $E=10^{9}\,{\rm eV}$ to $E=10^{15}\,{\rm eV}$. The output of these simulations was afterwards injected into a MARTA Geant4 dedicated simulation, leading to an estimate of the number of muons able to reach the RPC pads. The rate is obtained using the expected number of muons in the water-Cherenkov detector, taken from \cite{SDcalib}. From this study, one can conclude that the hits due to atmospheric particles are of the order of 5-7 Hz.

The results, presented in figure \ref{fig:fig7}, show that the number of atmospheric particles in each pad per minute is higher than 300, which means that a statistical precision of 1\% is reached every half an hour. 

\subsection{MARTA array}

Many possibilities arise from the knowledge of the arrival position and time of individual muons in MARTA stations. Some of the main aspects are outlined below. 

\vspace{2mm}
\noindent\textbf{From muon density to composition}

\begin{figure}
\resizebox{0.45\textwidth}{!}{%
  \includegraphics{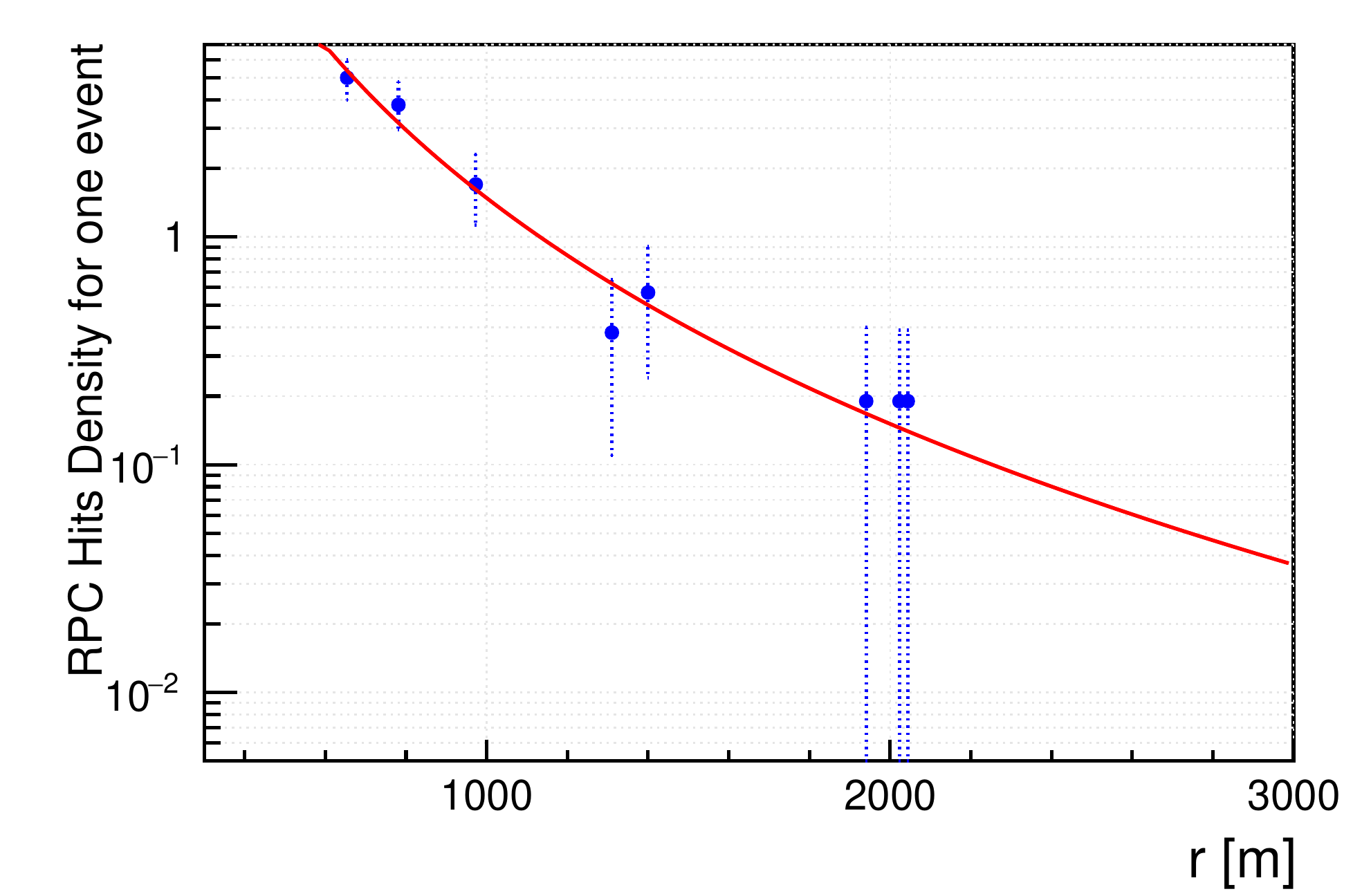}}
\caption{\label{fig:fig8} LDF of one QGSJET-II.04 proton shower with $E=10^{19.8}$ eV and $\theta$=38$^\circ$. RPC hits densitiy are given in square meters.}
\end{figure}

\noindent In figure \ref{fig:fig8} is shown the lateral distribution function (LDF) obtained by the MARTA RPCs for a simulated proton shower with $E=10^{19.8}$ eV and $\theta=38^\circ$ simulated with CORSIKA and QGSJET-II.04. The LDF, which represents the density of particle at a given distance from the shower core, was, in a first approximation, parametrized by \cite{auger}:

\begin{equation}
\rho_{LDF}(r,\beta) = \rho_{1000} \left( \frac{r}{1000} \right)^\beta \cdot \left( \frac{700+r}{1000+700} \right)^\beta   
\end{equation}

\noindent where the parameters $\rho_{1000}$ and $\beta$ represent, respectively, the normalization and the shape. In the fit, $\beta$ was fixed to -2.1.

\begin{figure}
\resizebox{0.45\textwidth}{!}{%
  \includegraphics{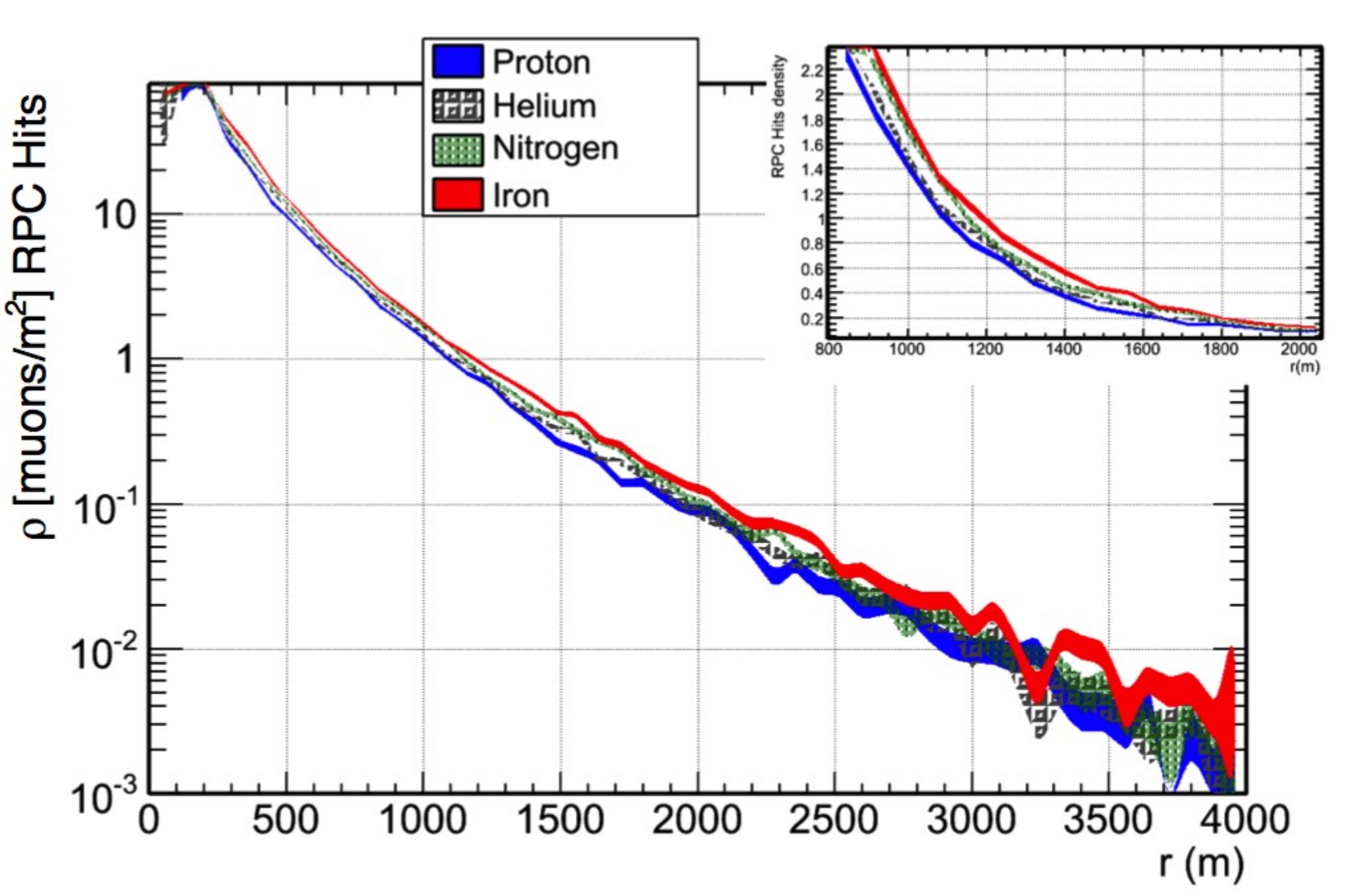}}
  \resizebox{0.45\textwidth}{!}{%
  \includegraphics{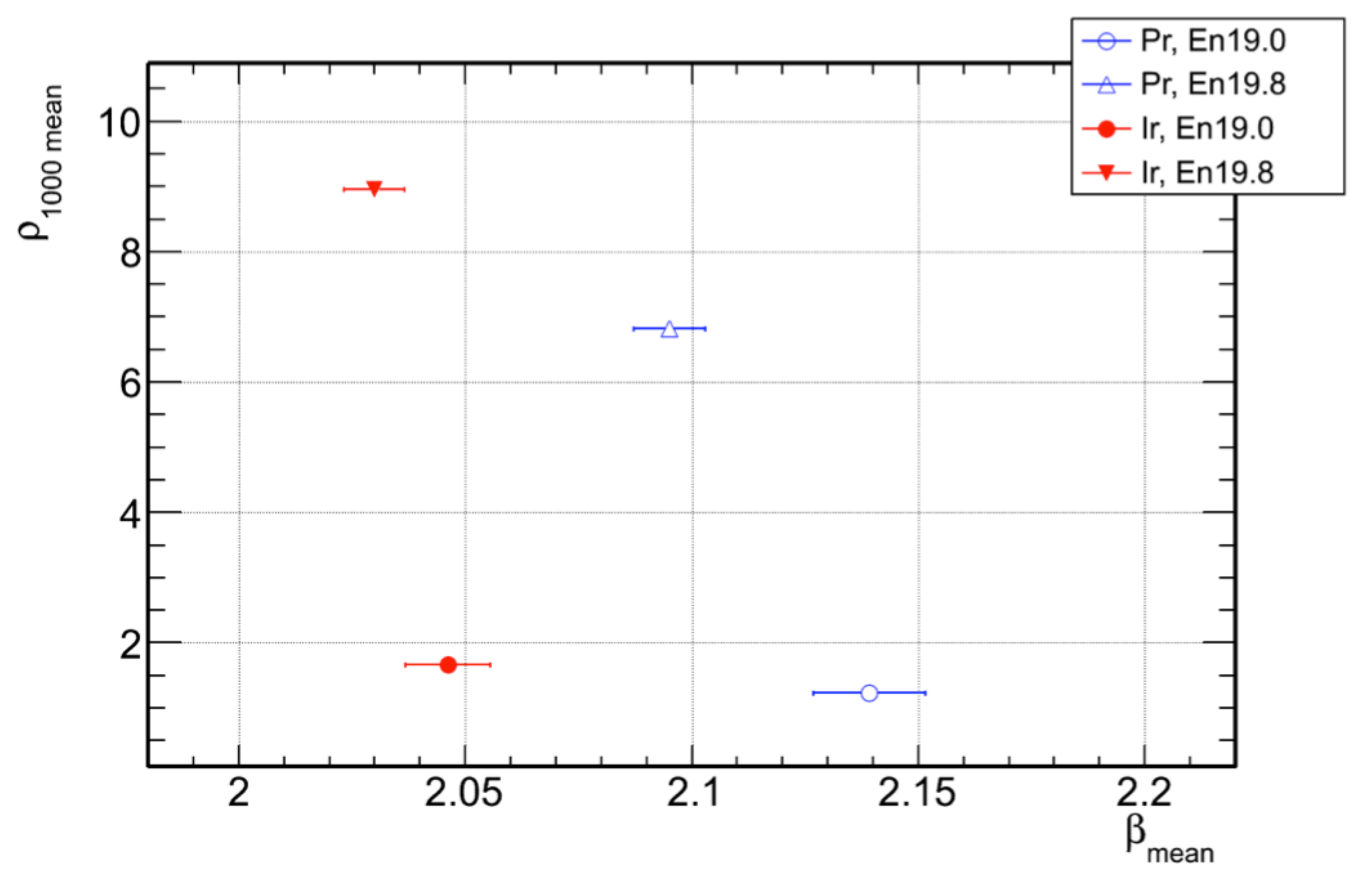}}
\caption{\label{fig:fig9} Top: Average MARTA LDF (over 300 events) for different primaries with QGSJET-II.04 at $E=10^{19}$ eV and $\theta=38^\circ$; Bottom: $\rho_{{1000}_{\rm mean}}$ and $\beta_{\rm mean}$.}
\end{figure}

In figure \ref{fig:fig9}(top) the average LDF for proton, helium, nitrogen and iron QGSJET-II.04 showers at $E=10^{19}$ eV and $\theta$=38 $^\circ$ is shown. This figure was obtained using 300 showers for each primary. A net separation is visible. Considering an energy bin of $\log(E/{\rm eV})=0.1$ around $E=10^{19}\,$eV, and taking the ultra-high-energy cosmic ray flux~\cite{augerSpectrum}, an experiment such as the Pierre Auger Observatory would be able to reach this event statistics in less than half a year.  
It is then possible to fit such average distributions taking as free parameters both $\rho_{1000}$ and $\beta$ ($\rho_{{1000}_{\rm mean}}$ and $\beta_{\rm mean}$). In figure \ref{fig:fig9}(bottom), $\rho_{{1000}_{\rm mean}}$ and $\beta_{\rm mean}$ for proton and iron primaries at $E=10^{19}$ eV and $E=10^{19.8}$ eV are shown. A clear separation between proton and iron is observed, showing that $\beta_{\rm mean}$ may be a powerful variable to assess the beam composition. The impact on the choice of the high-energy hadronic interaction model was evaluated repeating the analysis using EPOS-LHC instead of QGSJet-II.04. Although the results for $\beta$ and $\rho_{1000}$ obtained using EPOS-LHC are slightly higher than from QGSJet-II.04 (about 15\% for $\rho_{1000}$ and 2\% for $\beta$), the discrimination power regarding primary mass composition is not altered significantly.

\begin{figure}
\resizebox{0.45\textwidth}{!}{%
  \includegraphics{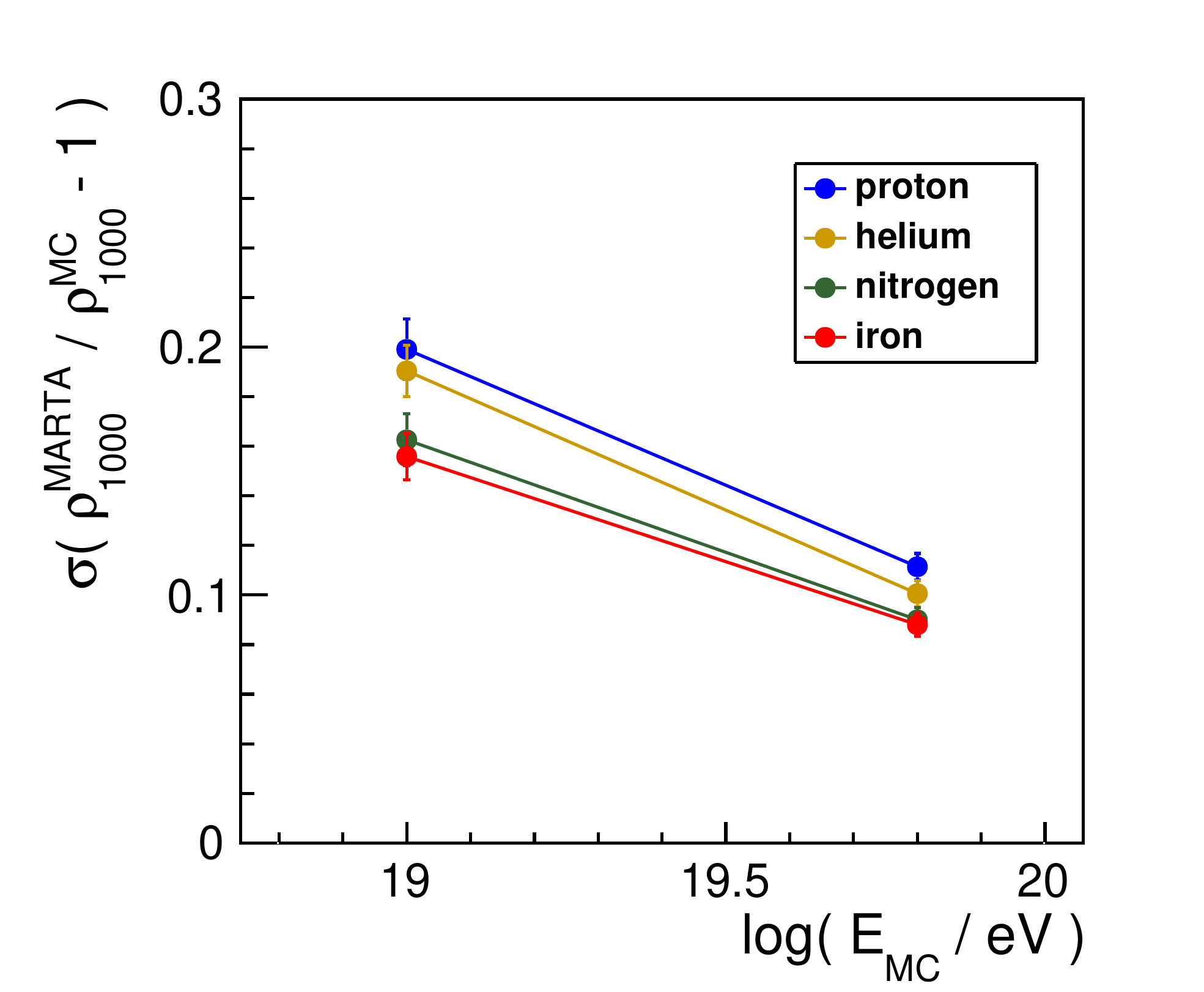}}
  \resizebox{0.45\textwidth}{!}{%
  \includegraphics{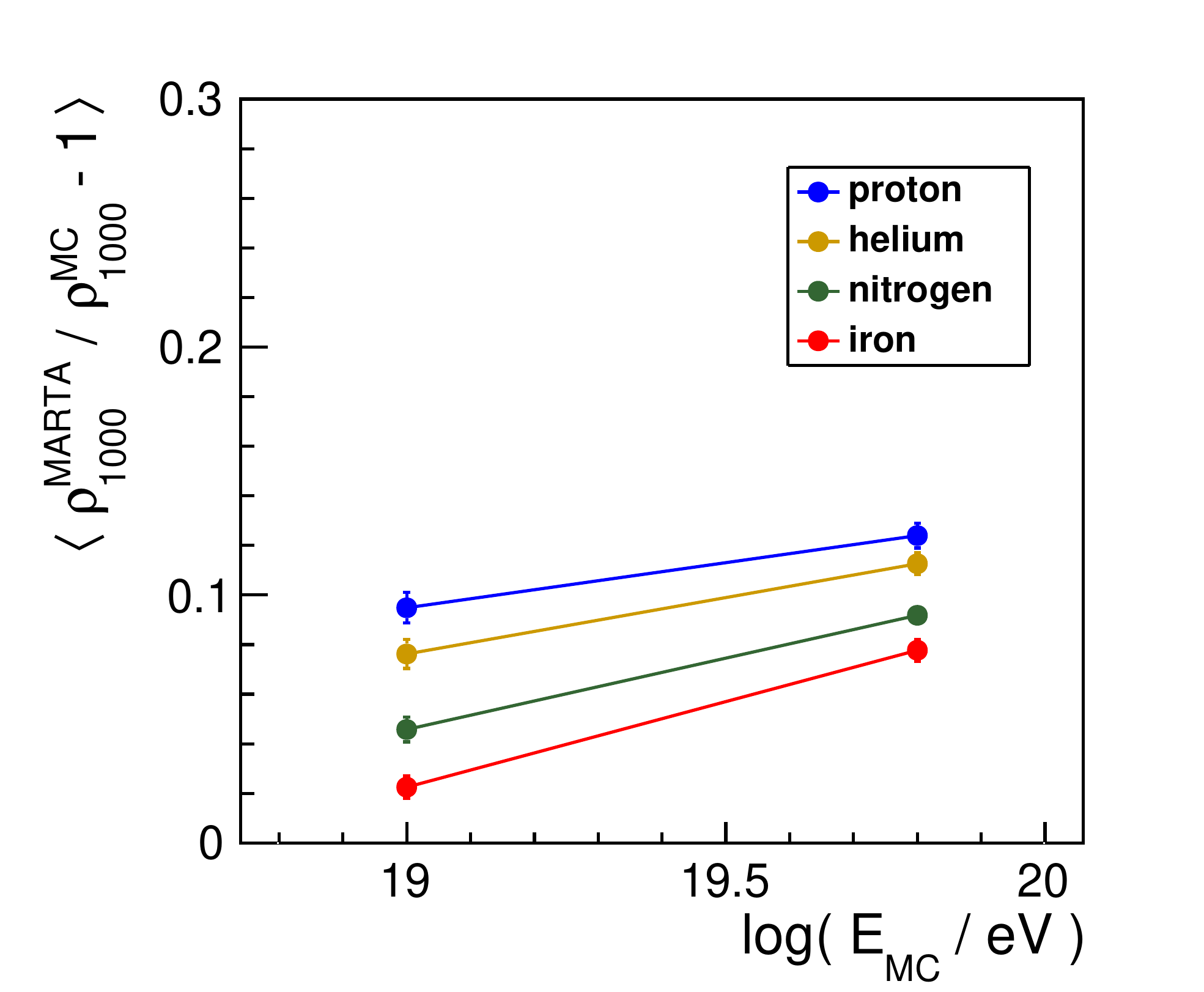}}
\caption{\label{fig:fig10} $\rho_{1000}^{\rm MARTA}$ resolution (top) and bias (bottom) for different primaries at $\theta=38^\circ$, for $E=10^{19}$ eV and  $E=10^{19.8}$ eV obtained with QGSJET-II.04. Similar results were obtained with the EPOS-LHC interaction model.} 
\end{figure}

\begin{figure*}
\centering
\resizebox{0.85\textwidth}{!}{%
  \includegraphics{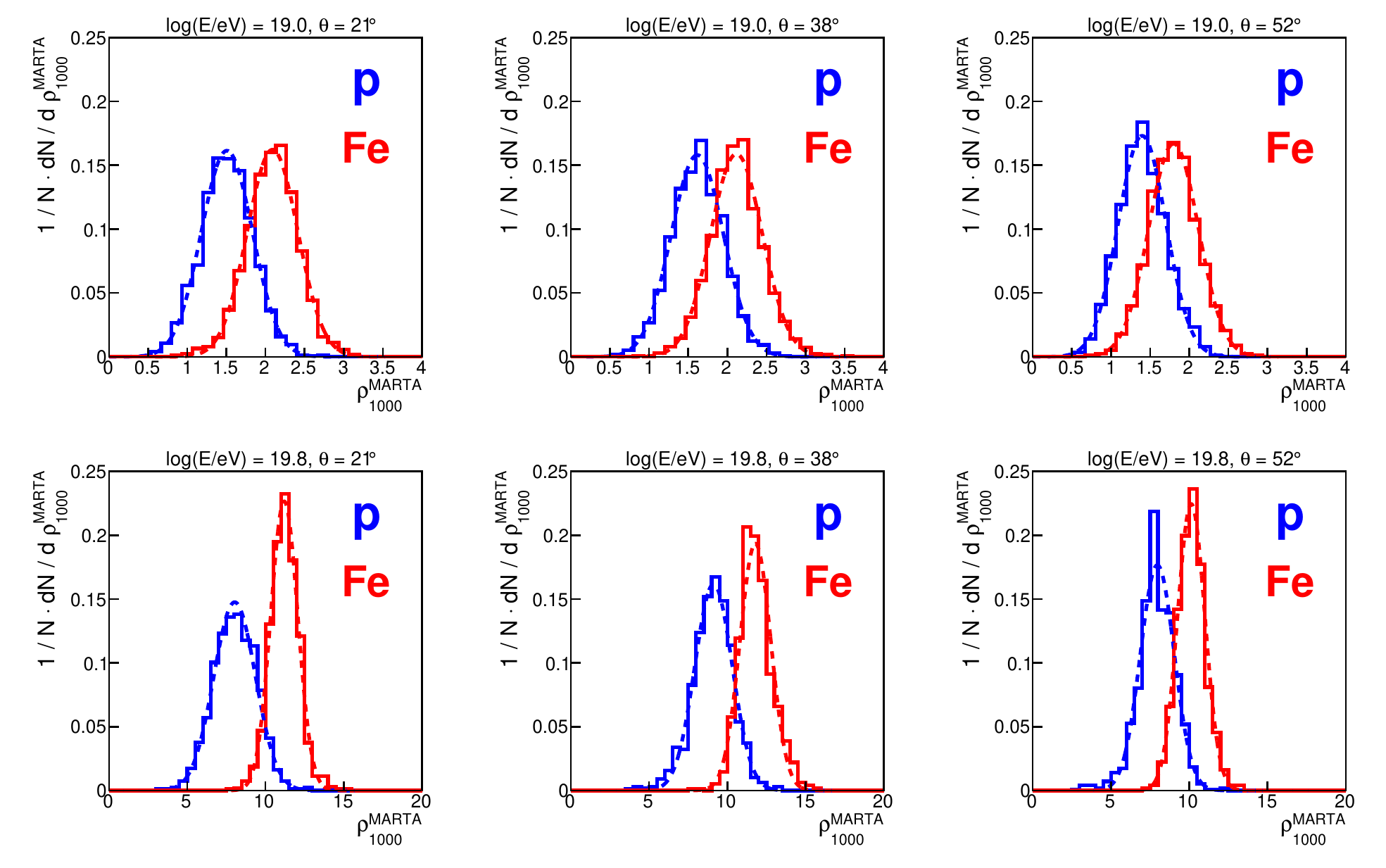}}
 \caption{\label{fig:fig11} $\rho_{1000}^{\rm MARTA}$ distributions for proton and iron primaries at different zenith angles (from left to right, $\theta=21^\circ, 38^\circ, 52^\circ$ with $E=10^{19}$ eV (top plots) and $E=10^{19.8}$ eV (bottom plots) obtained with QGSJET-II.04. Similar results were obtained with the EPOS-LHC interaction model.} 
\end{figure*}

Fitting the individual muon LDF distributions, and fixing $\beta$ as a function of the zenith angle, one obtains for each shower $\rho_{1000}$, the measured muon density at 1000 m from the core ($\rho_{1000}^{\rm MARTA}$). The $\beta$ parameter was obtained from a mixed composition simulation (50\% proton and 50\% iron). The bias and resolution with respect to the true muon density are shown in figure \ref{fig:fig10}, for different primary types at $\theta = 38^\circ$, for  $E=10^{19}$ eV and  $E=10^{19.8}$ eV. 
This figure shows that the bias in the muon density estimator $\rho_{1000}^{\rm MARTA}$ is nearly energy-independent. It has also been observed that the bias decreases with the increasing  zenith angle and has no significative dependence on the hadronic model considered.
In figure \ref{fig:fig11} we show the distributions of $\rho_{1000}^{\rm MARTA}$ for p and Fe at different angles and for two energy values.

\vspace{2mm}
\noindent\textbf{From muon production depth to composition}

\noindent In the muon production depth (MPD) technique, the shower geometry is combined with the arrival times of muons to reconstruct the longitudinal profile of muon production \cite{mpd1,mpd2}. While the reconstruction of the MPD is the same as for the Auger WCDs, the direct detection of muons opens great possibilities to extend the use of this technique.

\begin{figure}
\resizebox{0.45\textwidth}{!}{%
  \includegraphics{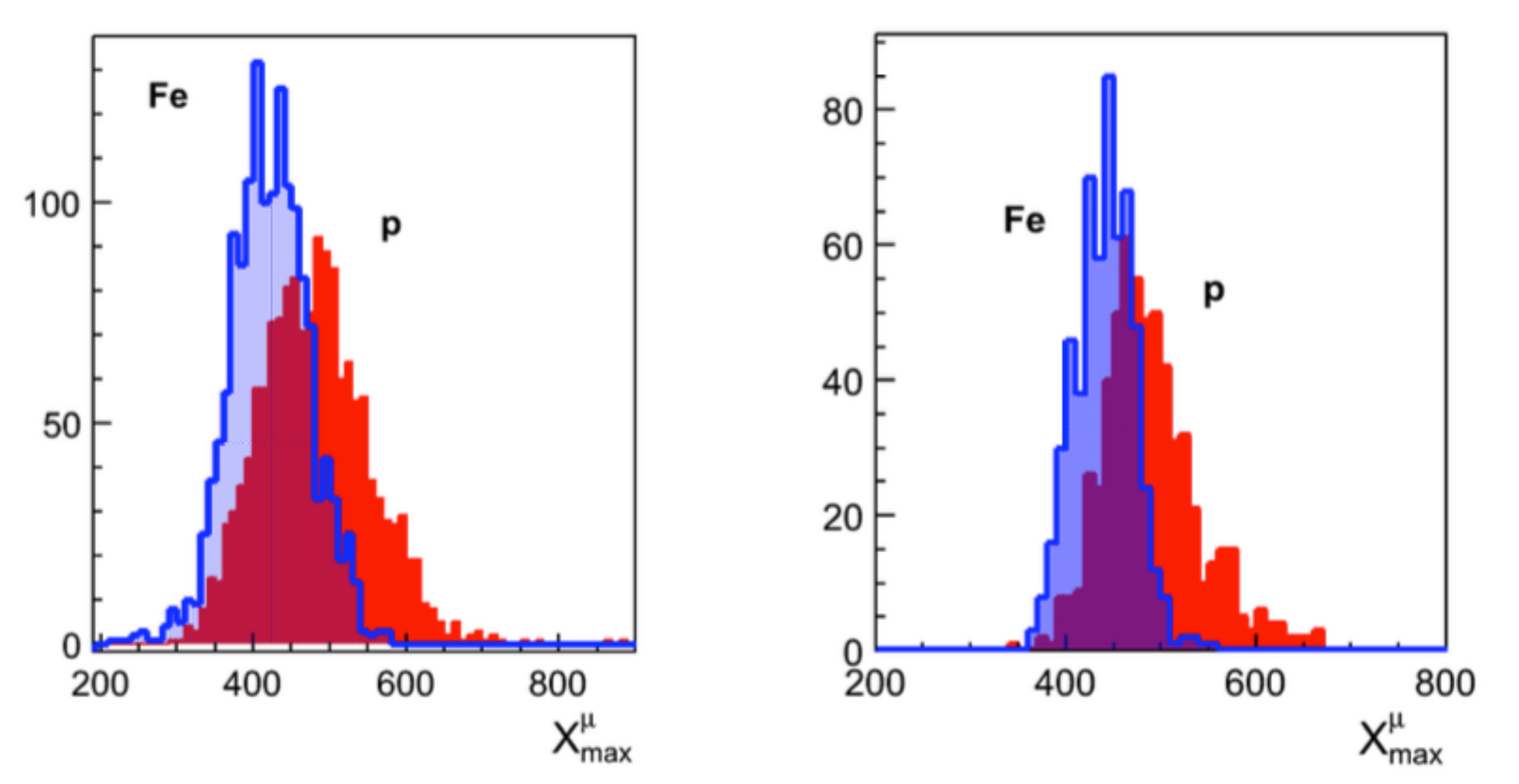}}
\caption{\label{fig:fig12} $X^\mu_{max}$ distributions for simulated showers initiated by protons and iron at $\theta=38^\circ$ with $E=10^{19}$ eV (left) and $E=10^{19.8}$ eV (right) obtained with QGSJET-II.04.} 
\end{figure}

The maximum of this profile is known to be composition-sensitive variable~\cite{MPDpheno} which could add relevant information about the hadronic interaction processes that rule the shower development. As an example, the depth of the maximum of the MPD for proton and iron showers is shown in figure \ref{fig:fig12}, for $\theta=38^\circ$ and two different primary energies. The event resolution stays the same as it is dominated by the number of muons entering the reconstruction, which will remain practically the same in each station. The largest impact comes from the reduction of systematics which can be lowered by $\sim 10{\rm g/cm^2}$, since the uncertainties associated to the time response of the WCD are reduced by the RPCs to negligible values (compared to the remaining sources of systematics). In addition, the  separation of the electromagnetic and muonic shower components at ground allows the MPD technique to be applied to stations which are closer to the core and also in more vertical showers, which permits the enlargement of the energy window and angular window of applicability. Finally, it is worth noting that the angular dependence of MPD can give information about the EAS muon energy spectrum~\cite{EmuEAS}, providing an additional handle to evaluate and constrain hadronic interaction models.

\begin{figure*}
\centering
\resizebox{0.75\textwidth}{!}{%
  \includegraphics{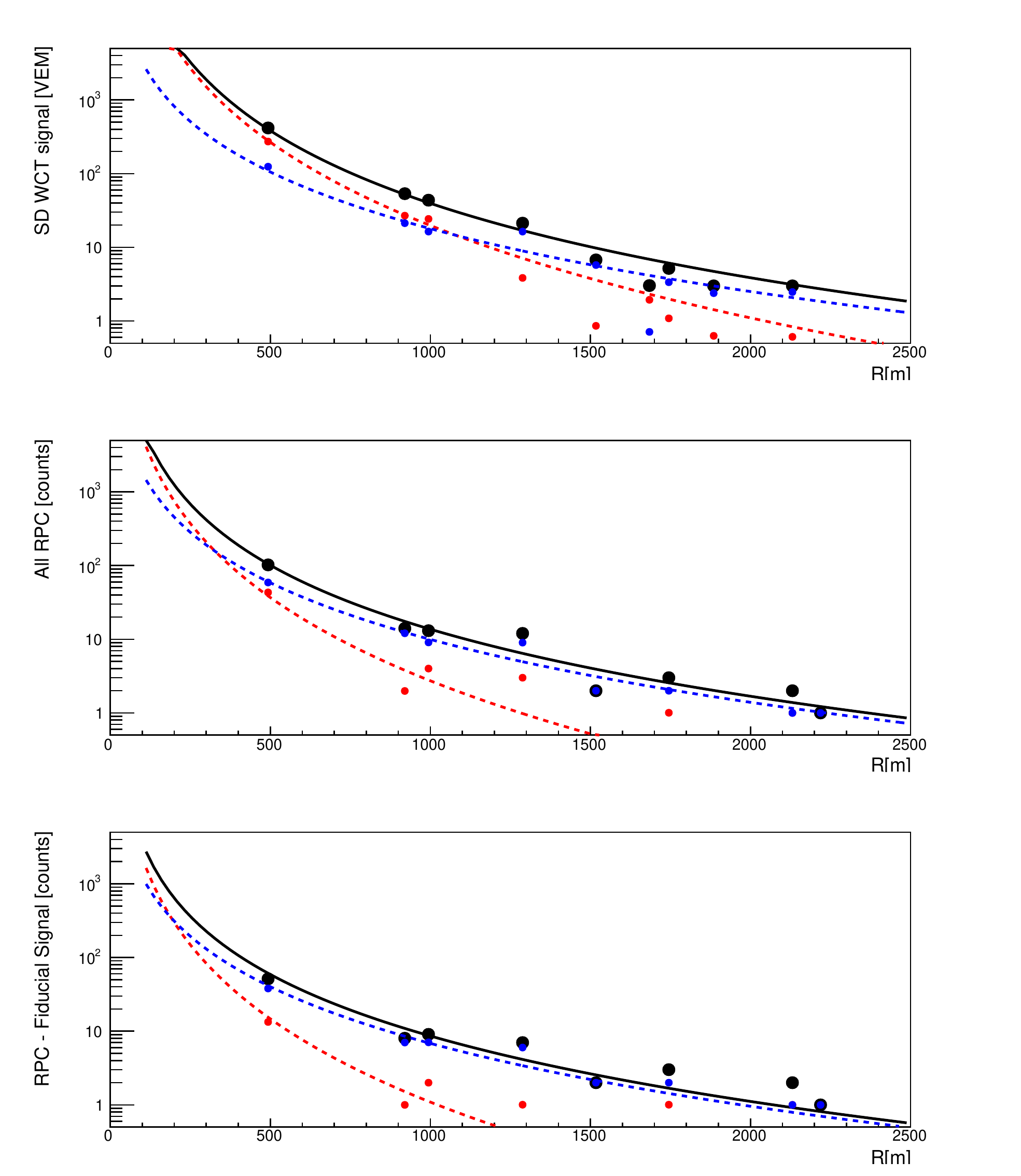}}
\caption{\label{fig:fig13}  Combined fit for one event (QGSJET-II.04 proton with $\theta=38^\circ$ and $E=10^{19}$ eV) in the tank  and in the RPC's total and fiducial area. The large black dots are the measurements used in the fit. The black line is the sum of the two dashed lines corresponding to the muonic (blue) and electromagnetic (red) LDFs. The true (simulated) information about the muonic and electromagnetic signals is shown as smaller dots.} 
\end{figure*}

\subsection{Combined measurements}

The redundant use of the two sub-detectors, detecting the same portion of the shower at consecutive depths, opens several possibilities for systematic studies and cross-calibration, reducing global uncertainties associated with the measurements. Additionally, there is the possibility to explore new variables and increase the amount of reconstructed information by cross-analyses between the two sub-detectors, benefiting for example the precision of the energy spectrum measurement or the sensitivity to photon primaries, which have a much-reduced muon content compared to hadron primaries~\cite{easPhoton}.

While a detailed exploitation of such possibilities is out of the scope of this concept paper, we give as an example the impact on the energy reconstruction of correlating the information of the two detectors to disentangle the muonic and electromagnetic shower components. The electromagnetic and muonic contents of the shower can be estimated performing combined fits to the LDFs measured both in the RPC fiducial and non-fiducial areas as well as in the water-Cherenkov detector. The fitted parameters $S_{em}$ and $S_{\mu}$ should be proportional, respectively, to the total electromagnetic energy and the total number of muons in the shower. In figure \ref{fig:fig13} an example of such a fit is shown. 

Taking advantage of the excellent timing of RPCs, the trajectory of single EAS muons can be reconstructed using an MPD-like algorithm. The RPC segmentation allows one to determine the path of the muon through the SD tank, and the track length can be reconstructed with a resolution of around 6\%. With this information, one can compare the expected muon signal in the WCD with the collected one, constraining in this way either the WCD properties or the muon energy spectrum. The usual WCD calibration with atmospheric muons can be improved requiring a coincidence with MARTA RPCs. These studies, combined with the data collected by an RPC hodoscope installed in a test WCD at the Observatory site can help us understand and improve the monitoring tools for each water-Cherenkov detector in the field equipped with a MARTA unit \cite{gianni}.

\section{Summary and prospects}

The MARTA concept for the direct measurement of muons in air showers has been presented. An RPC-based detector measures muons with very good space and time resolution and is shielded by a detector able to absorb and measure the electromagnetic component of showers. The combination of the two detectors in a single, compact detector unit provides a unique measurement that opens rich possibilities in the study of air showers.

MARTA was designed to fulfil the requirements of large high-energy cosmic-ray experiments, proposing low cost units able to operate stably and at high efficiency in outdoor environments with minimal maintenance and low power consumption. To provide a concrete and realistic performance expectations a detailed implementation for the Pierre Auger Observatory has been developed. 

Extensive R\&D on RPCs for MARTA was carried out. Long-term performance tests confirm  stable operation with $\sim$90\% efficiency and low gas flow ($0.4\, {\rm cm^3/min}$). About twenty MARTA RPC units are installed and taking data in the laboratory and at the field at the Pierre Auger Observatory. A MARTA mini-array, equipping eight water-Cherenkov detectors with four RPC modules each, will be installed at the Pierre Auger Observatory site. Each MARTA station will count muons with a time resolution of 5 ns and a position resolution limited by the pad dimensions. In each individual station, the number of muons will be reconstructed with a bias and resolution below 20\% in a wide range of distances to the shower core. 

In a MARTA array, many possibilities arise from the knowledge of the arrival position and arrival time of muons. In particular, composition sensitivity is greatly enhanced. Simulation studies show that  the lateral density of muons at 1000 m from the core is measured with a resolution of 10-20\% and a bias below 10\%, and that differences in the lateral distribution of muons provide a clear separation between proton and iron.
In addition, the MPD technique is applicable to a wider range of zenith angles, and the associated systematics greatly benefit from the excellent time resolution. The use of two systems detecting the same portion of the shower at consecutive depths opens several possibilities for systematic studies, cross-calibrations and cross-analyses, benefiting, for example, the sensitivity to photon primaries.

The MARTA concept for the direct and precise measurement of muons in air showers may find application in future cosmic-ray experiments, but also in dedicated arrays for high-energy gamma rays. In this case, RPCs provide very good space and time resolution, while the calorimetric detector ensures trigger efficiency and background rejection. Combined, they can bring the sensitivity of air shower gamma-ray arrays to lower energies \cite{lattes}. While the concept applies, developments concerning operation at  very high altitudes will have to be addressed.

\section*{Acknowledgements}

The authors greatly acknowledge all the fruitful discussions held with Pierre Auger colleagues. We are also very thankful for all the support at the site and the use of Pierre Auger Observatory infrastructure, in particular to the Observatory staff for the assistance and dedication.
The authors also thankfully acknowledge the support of FAPESP/FCT (2014/19946-4). S. Andringa, L. Cazon, R. Concei\c{c}\~ao, R. Luz, R. Sarmento want to thank funding by Funda\c{c}\~ao para a Ci\^encia e Tecnologia (PD/ BD/ 113488 /2015, SFRH/ BPD/ 84304/ 2012). J. \u{R}\'{i}dk\'y, P. Tr\'avn\'{i}\u{c}ek, J. V\'{i}cha acknowledge the financial support under Grant No. MSMT CR LTT18004, LM2015038 and CZ.02.1.01/ 0.0/ 0.0/ 16\_013/ 0001402.

%
%

%
%


\bibliographystyle{spphys.bst}
\bibliography{BibMARTA}

\end{document}